\documentclass[
reprint,
superscriptaddress,
showpacs,preprintnumbers,
nofootinbib,
amsmath,amssymb,
aps,
prl,
twocolumn
]{revtex4}

\usepackage{graphicx,color}

\usepackage{textcomp}
\usepackage{graphicx}
\usepackage{dcolumn}
\usepackage{bm}
\usepackage{hyperref}

\usepackage{color}

\begin{document}

\preprint{Bombor}

\title{Half-Metallic Ferromagnetism in the Heusler Compound Co$_2$FeSi revealed by Resistivity, Magnetoresistance, and Anomalous Hall Effect measurements}

\author{Dirk~Bombor}
\email{d.bombor@ifw-dresden.de}
\affiliation{IFW Dresden, 01171 Dresden, Germany}
\author{Christian~G.~F.~Blum}
\affiliation{IFW Dresden, 01171 Dresden, Germany}
\author{Oleg~Volkonskiy}
\affiliation{IFW Dresden, 01171 Dresden, Germany}
\author{Steven~Rodan}
\affiliation{IFW Dresden, 01171 Dresden, Germany}
\author{Sabine~Wurmehl}
\affiliation{IFW Dresden, 01171 Dresden, Germany}
\affiliation{Institut f\"ur Festk\"orperphysik, Technische Universit\"at Dresden, D-01062 Dresden, Germany}
\author{Christian~Hess}
\email{c.hess@ifw-dresden.de}
\affiliation{IFW Dresden, 01171 Dresden, Germany}
\author{Bernd~B\"uchner}
\affiliation{IFW Dresden, 01171 Dresden, Germany}
\affiliation{Institut f\"ur Festk\"orperphysik, Technische Universit\"at Dresden, D-01062 Dresden, Germany}

\date{\today}

\begin{abstract}
	We present electrical transport data for single-crystalline Co$_2$FeSi which provide clear-cut evidence that this Heusler compound is truly a half-metallic ferromagnet, i.e. it possesses perfect spin-polarization. More specifically, the temperature dependence of $\rho$ is governed by electron scattering off magnons which are thermally excited over a sizeable gap $\Delta\approx 100\,\text{K}$ ($\sim 9\,\text{meV}$) separating the electronic majority states at the Fermi level from the unoccupied minority states. As a consequence, electron-magnon scattering is only relevant at $T\gtrsim\Delta$ but freezes out at lower temperatures, i.e., the spin-polarization of the electrons at the Fermi level remains practically perfect for $T\lesssim\Delta$. The gapped magnon population has a  decisive influence on the magnetoresistance and the anomalous Hall effect (AHE): i) The magnetoresistance changes its sign at $T\sim 100\,\text{K}$, ii) the anomalous Hall coefficient is strongly temperature dependent at $T\gtrsim 100\,\text{K}$ and compatible with Berry phase related and/or side-jump electronic deflection, whereas it is practically temperature-independent at lower temperatures.
\end{abstract}

\maketitle

\begin{figure}
		\includegraphics{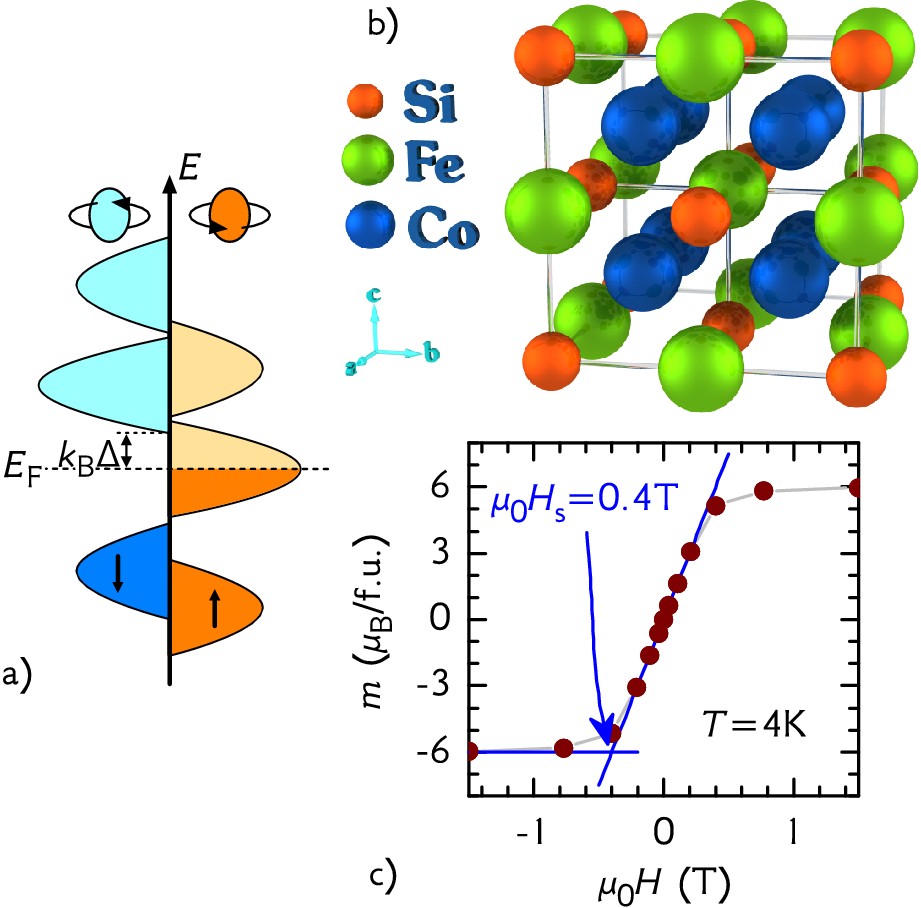} \centering
		\caption[]{a) L2$_1$ crystal structure of Co$_2$FeSi. b) Sketch of the density of states of a HMF in dependence of energy for minority and majority electrons with energy gap at Fermi level $E_\text{F}$ for minority electrons. c) Magnetic moment of Co$_2$FeSi single crystal.}
		\label{fig:intro}
	\end{figure}

	The class of Heusler compounds offers a tool box for studying various materials funtionalities and interesting physical ground states \cite{Chadov10,Graf11,Liu12}. For example, for some Co$_2$-based Heusler compounds full spin polarization at the Fermi level referred to as half-metallic ferromagnetism \cite{degroot1983,mueller2009,shan2009,Graf11} has been predicted [see fig.~\ref{fig:intro}a for an illustration of the density of states (DOS) of a half-metallic ferromagnet (HMF)], which renders these materials prime candidates for spintronics applications. Most prominent among these materials is the Heusler compound Co$_2$FeSi, since this ferromagnetic metal possesses the highest Curie temperature ever found in Heusler compounds ($T_\text{C}=1100\,\text{K}$) together with a large magnetic moment of $m=6\,\mu_\text{B}/\text{f.u.}$ \cite{wurmehl2005}.
	
	In this letter we present the electronic transport properties, viz., resistivity, magnetoresistance and Hall effect, of high quality Co$_2$FeSi single crystals \cite{blum}. Our data provide fresh evidence of Co$_2$FeSi being a true HMF: At high temperatures ($T\gtrsim 100\,\text{K}$), the resistivity $\rho$ is governed by scattering off ferromagnetic magnons. This magnon scattering process is exponentially suppressed at lower temperatures which implies a complete suppression of the minority electron density of states at the Fermi level where the excitation gap for these levels is extracted from resistivity as $\Delta\sim100\,\text{K}$. This crucial finding is further corrobarated by a sign change of magnetoresistance from negative at high temperatures ($T \gtrsim 110 \,\text{K}$) to positive at lower temperatures. Furthermore, we observe that the activation of ferromagnetic magnons across $\Delta$ has a profound impact on the anomalous Hall effect (AHE). While the anomalous Hall coefficient is practically temperature independent at $T \lesssim 100 \,\text{K}$, consistent with dominating extrinsic skew scattering or the ordinary Lorentz-force induced Hall effect, it strongly increases at higher $T$, where it follows $\sim\rho^2$, suggestive of dominating Berry phase and/or side-jump contributions caused by activated ferromagnetic magnons.
	
	\textit{Experimental details} -- Large single crystals of Co$_2$FeSi, which crystallizes in the L2$_1$ structure with a unit cell size of $a=5.64\,\text{\AA}$ \cite{wurmehl2006} (fig.~\ref{fig:intro}b), were prepared using the floating zone technique \cite{blum}. Magnetization measurements confirm the predicted magnetization of $m=6\,\mu_\text{B}/\text{f.u.}$ and yield a saturation field of $\mu_o H_\text{s}=0.4\,\text{T}$ (see fig.~\ref{fig:intro}c). The susceptibility below $H_\text{s}$ is $\chi=5.4$. Due to the high Curie temperature, the temperature dependences of the saturation field and susceptibility are negligible up to room temperature. For resistivity measurements in magnetic field, a single crystal with dimensions of $a=b=c=3\,\text{mm}$ was used. Resistivity was measured using a standard 4-probe alternating current dc-technique. Hall measurements were done simultaneously using 2 additional Hall voltage contacts.
	
\begin{figure}
		\includegraphics{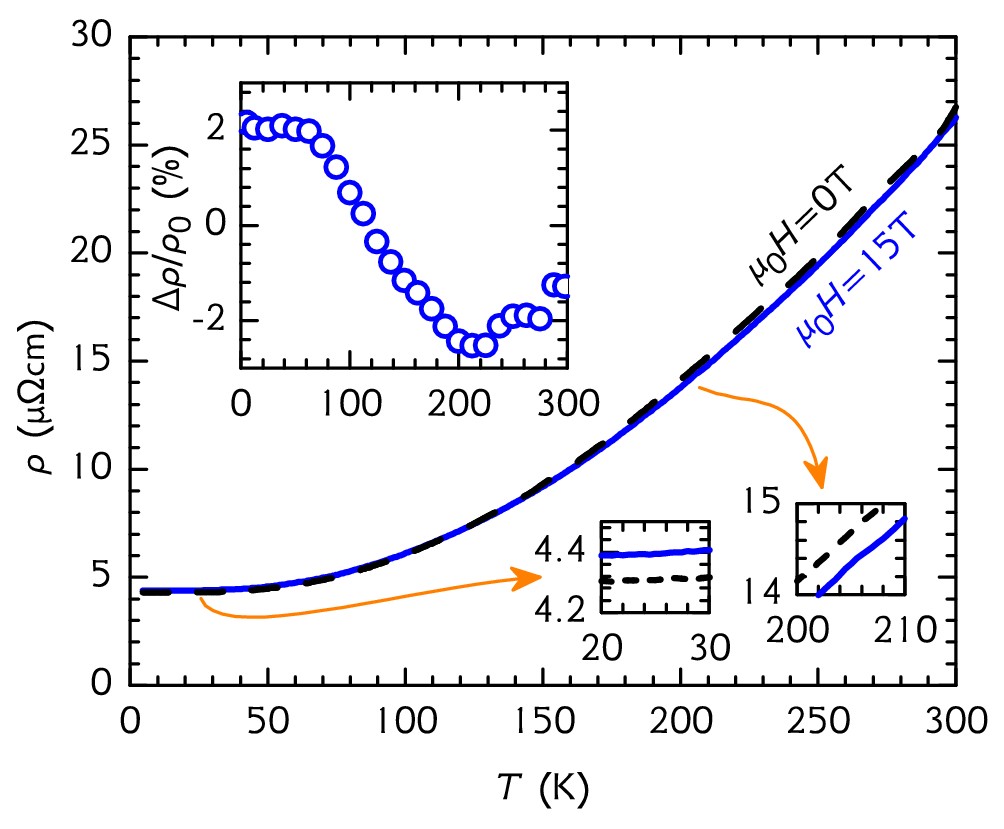} \centering
		\caption[$\rho(T)$ in magnetic field]{Temperature dependence of the resistivity in zero and $15\,\text{T}$ magnetic field. Upper left inset: Difference ($\Delta\rho=\rho(15\text{T})-\rho(0\text{T})$, $\rho_0=\rho(0\text{T})$) of these measurements. Lower right insets: Close-up to low and high temperatures.}
		\label{fig:rhotb}
	\end{figure}

	\textit{Resistivity} -- Fig.~\ref{fig:rhotb} shows the resistivity $\rho$ as a function of temperature $T$ for zero magnetic field and for $\mu_0 H = 15 \,\text{T}$. In zero field the resistivity shows a clear metallic temperature dependence with a monotonic increase with increasing temperature. A residual resistivity	of $\rho _ \text{R}\approx4\,\text\textmu\Omega\text{cm}$ and a residual resistivity ratio of $\rho(300\text{K})/\rho_\text{R}=6.5$ shows the extraordinary quality of our samples.
	
	The resistivity of metals is usually dominated by electron-phonon scattering. This typically leads to a linear temperature dependence at elevated temperatures ($T\gtrsim 50 \,\text{K}$) and a saturation towards residual resistivity at lower temperatures. Such a linear part is completely absent in Co$_2$FeSi. However, for ferromagnetic materials like the compound under scrutiny, another scattering process may be found, i.e., electron-magnon scattering, which often leads to a quadratic temperature dependence of the resistivity (see e.g.\ \cite{goodings1963,isshiki1978,isshiki1984,raquet2002,raquet2002a,otto1989}).
	
\begin{figure}
		\includegraphics{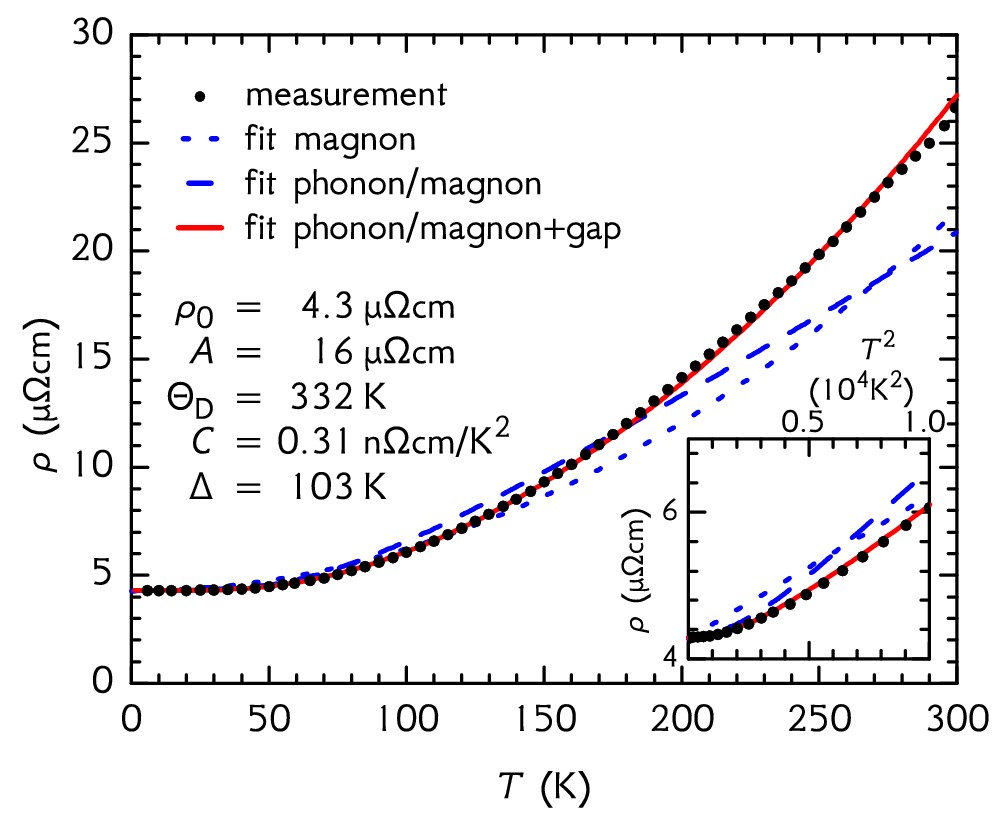} \centering
		\caption[$\rho(T)$]{Resistivity in dependence of the temperature. Measured values are displayed as black dots. Blue chopped lines show fits with magnonic ($\sim T^2$) and magnonic/phononic temperature dependence. The red continuous line shows a fit where an additional Boltzmann factor was added to the magnonic scattering, its fitting parameters are displayed (see text for formulas).}
		\label{fig:rhofit}
	\end{figure}

	However, the attempt to fit $\rho(T)$ of Co$_2$FeSi with a quadratic and a quadratic plus linear temperature dependence (for pure magnon and magnon/phonon scattering of the electrons, respectively) fails completely (see fig.~\ref{fig:rhofit}). Thus, the conventional electron-magnon scattering of band ferromagnets cannot account for the observed temperature dependence of the resitivity. It can, however, be rationalized much better if the predicted half-metallic character of the material is taken into account. In HMF, the minority spin electronic states are completely gapped at the Fermi level, where $k_\text{B}\Delta$ is the minimum excitation energy of majority charge carriers to occupy empty minority states involving a spin flip (see fig.~\ref{fig:intro}b). As a consequence, at low temperatures ($T \lesssim \Delta$) the generation of magnons (which involves a spin flip) and in particular electron-magnon scattering must be exponentially supressed. At higher temperatures ($T \gtrsim \Delta$) spin flips and magnon scattering would become possible when the Fermi distribution smears out the occupation of energy states around the Fermi level.
	
	In order to test against this scenario we fitted the resistivity with combined scattering of a residual ($\rho_\text{R}$), magnonic ($\rho_\text{M}$) and phononic ($\rho_\text{P}$) contribution:
	\begin{equation}
		\rho(T) = \rho_\text{R} + \rho_\text{M}(T) + \rho_\text{P}(T) \;.
	\end{equation}
	The residual resistivity $\rho_\text{R}$ is temperature independent and caused by defects of the ideal crystal lattice. The usually quadratic magnonic term ($\rho_\text{M}$) was weighted by a Boltzmann factor:
	\begin{equation}
		\rho_\text{M}(T) = C T^2 \cdot \text{e}^{-\Delta/T} \;.
	\end{equation}
	The parameter $C$ is a measure of the strength of the magnon scattering process. The phononic part of the scattering process ($\rho_\text{P}$) is described by the Bloch-Gr\"uneisen formula:
	\begin{equation}
		\rho_\text{P}(T) = A \left( \frac{T}{\Theta_\text{D}} \right)^5 \int_0^{\Theta_\text{D}/T} \frac{x^5}{(\text{e}^x-1)(1-\text{e}^{-x})} \text{d}x \;.
	\end{equation}
	With the Debye temperature $\Theta_\text{D}=332\,\text{K}$, determined from specific heat measurements, the resistivity fit \footnote{Note that electron-phonon scattering can be left out if one fits the data only up to $200\,\text{K}$.} yields an energy gap of
	\begin{equation}
		\Delta = 103 \,\text{K} \; , \;(k_\text{B}\Delta = 8.9 \,\text{meV}) \;\footnote{We estimate the uncertainty to less than $30\,\%$.}.
	\end{equation}
	 
	It is worth to point out that the substantial size of the gap implies that the spin-polarization of the electrons at the Fermi level remains practically perfect for $T\lesssim\Delta$.
	
	\begin{figure}
		\includegraphics{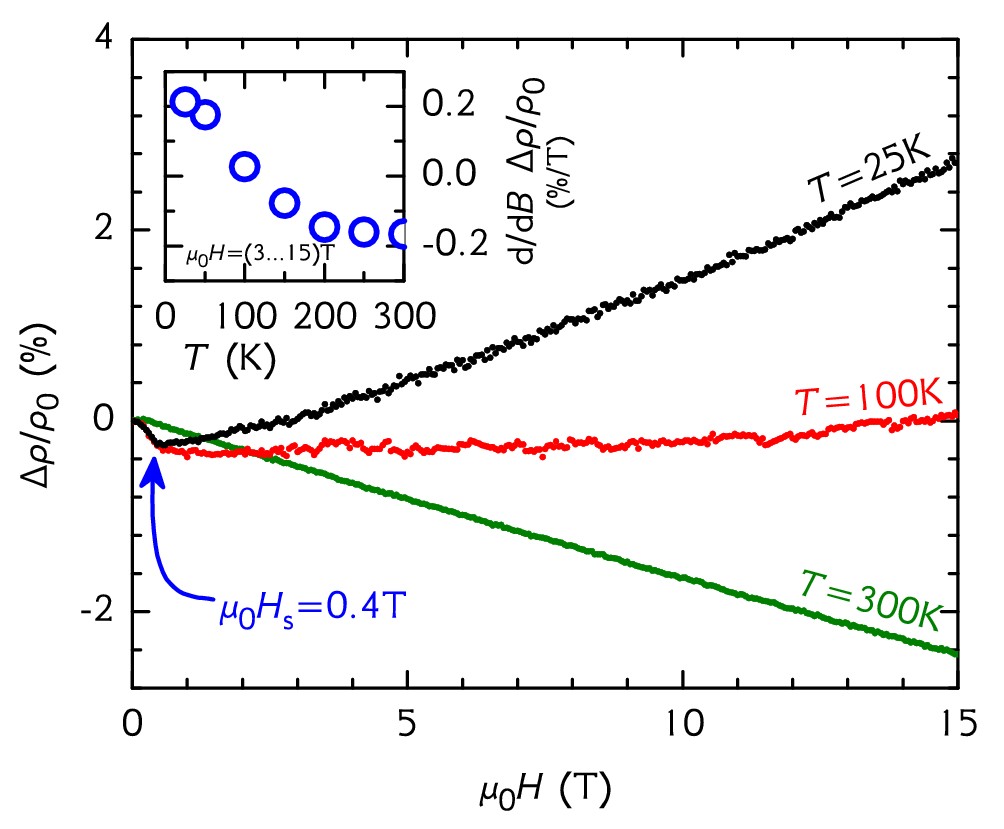} \centering
		\caption[$\rho(B)$ for different temperatures]{Magnetic field dependence of resitivity at different temperatures. Inset: slope of $\rho(B)$ in the applied field range of $\mu_0 H=(3\ldots 15)\,\text{T}$.}
		\label{fig:rhob}
	\end{figure}
	\textit{Magnetoresistance} -- An independent verification of the afore derived scenario of gapped-out magnon scattering can be obtained from the corresponding magnetoresistance of Co$_2$FeSi, because the influence of a magnetic field on the magnons should be reflected by the resistivity. As can be seen in fig.~\ref{fig:rhotb}, a magnetic field of $15\,\text{T}$ yields a relatively small ($\sim 2\,\text{\%}$) but well resolvable change of the resistivity as compared to the zero-field data. The inset highlights the sign and the temperature dependence of $\Delta\rho=\rho(15\text{T})-\rho(0)$. Apparently $\Delta\rho>0$ ($\Delta\rho<0$) for $T\lesssim 110\,\text{K}$ ($T\gtrsim 110\,\text{K}$).
	
	To get a more detailed view of this dependence on the magnetic field we measured the resistivity while sweeping the magnetic field and maintaining the temperatures (fig.~\ref{fig:rhob}). The inset of fig.~\ref{fig:rhob} shows the slope of the resistivity for fields higher than the saturation field. Up to the saturation field of $\mu_0 H_\text{s}=0.4\,\text{T}$ a small decrease of the resistivity can be seen, which can naturally be explained by alignment and joining of magnetic domains which decreases the probability of domain wall scattering. For high fields $H > H_\text{s}$, the resistivity decreases with increasing field at high temperatures; at low temperatures it increases with increasing field.
	
	This temperature dependence of the magnetoresistance can perfectly be explained by the afore presented scenario of the resistivity being dominated by scattering of spin-polarized electrons off gapped magnons. The negative magnetoresistance at high temperatures signals a decrease of the electronic scattering with increasing field. This is consistent with ferromagnetic magnons being the main scattering centers for electrons since a magnetic field enlarges the magnon energy and thus lowers the number of magnons. Upon lowering the temperature, magnons in a HMF are expected to freeze out exponentially since only one spin state is present at the Fermi level. Hence electron-defect scattering must become important at lower temperatures and one expects a more conventional positive magnetoresistance. Indeed, such a behaviour is observed at $T\lesssim 110\,\text{K}$: $\Delta\rho>0$ and $\Delta\rho(B)$ approaches a positive quadratic field dependence, which is characteristic for conventional multi-band metals.
	
\begin{figure}
		\includegraphics{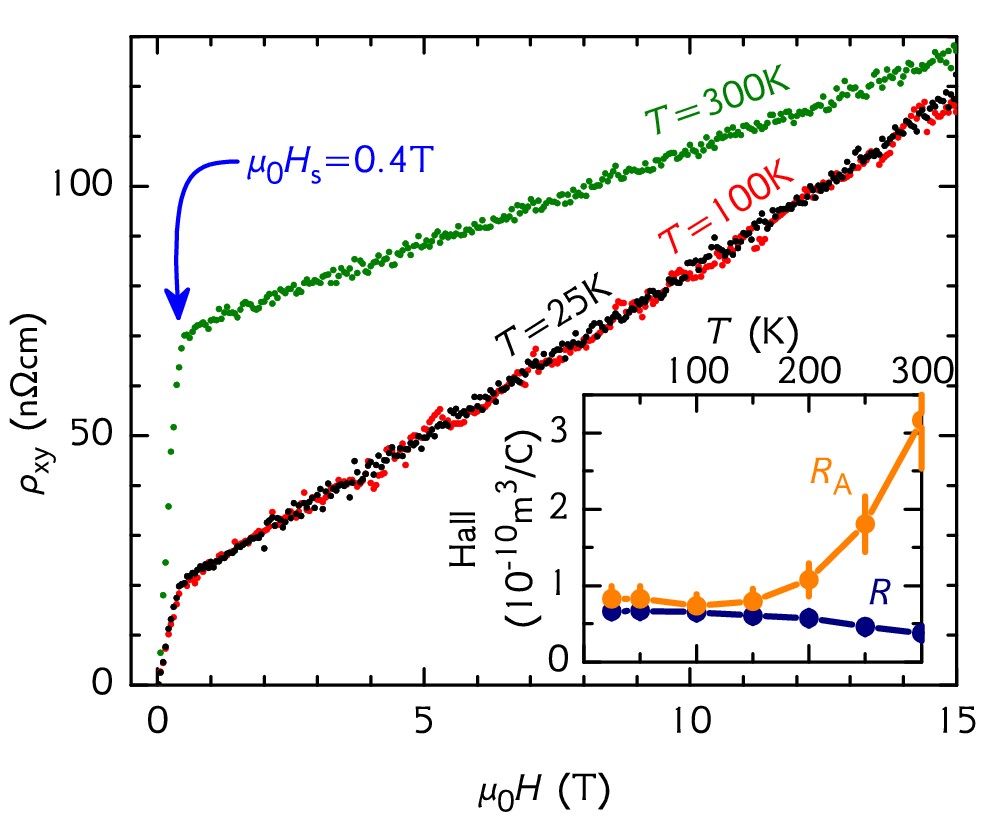} \centering
		\caption[$\rho_{xy}(B)$ for different temperatures]{Hall resistivity in dependence of the applied magnetic field for different temperatures. Inset: ordinary Hall coefficient $R$ and anomalous Hall coefficient $R_\text{A}$ in dependence of the temperature. }
		\label{fig:hall}
	\end{figure}

	\textit{Hall effect} -- Having established Co$_2$FeSi being a HMF and activated magnon generation for $T\gtrsim 100\,\text{K}$, we now move on to the anomalous Hall effect (AHE) of Co$_2$FeSi. The Hall resistivity of this compound is shown in fig.~\ref{fig:hall} along with its corresponding Hall coefficients in the inset. The Hall resistivity $\rho_{xy}(\mu_0 H)$ is characteristic for the AHE in ferromagnetic materials: it has a linear dependence on the applied magnetic field but with a strong kink and a changing slope just at the saturation field of $\mu_0 H_\text{s}=0.4\,\text{T}$. Since the Hall effect depends on the magnetic field $H$ as well as on its magnetization $M$, one considers two Hall coefficients, viz. the ordinary Hall coefficient $R$ and the so-called anomalous Hall coefficient  $R_\text{A}$, which connects the applied magnetic field $H$ and the magnetization $M$ of the sample, respectively, with the Hall resistivity \cite{otto1989,nagaosa2010}:
	\begin{equation}
		\rho_{xy} = \mu_0 \left( R \cdot H + R_\text{A} \cdot M \right) \;.
	\end{equation}
	
\begin{figure}
		\includegraphics{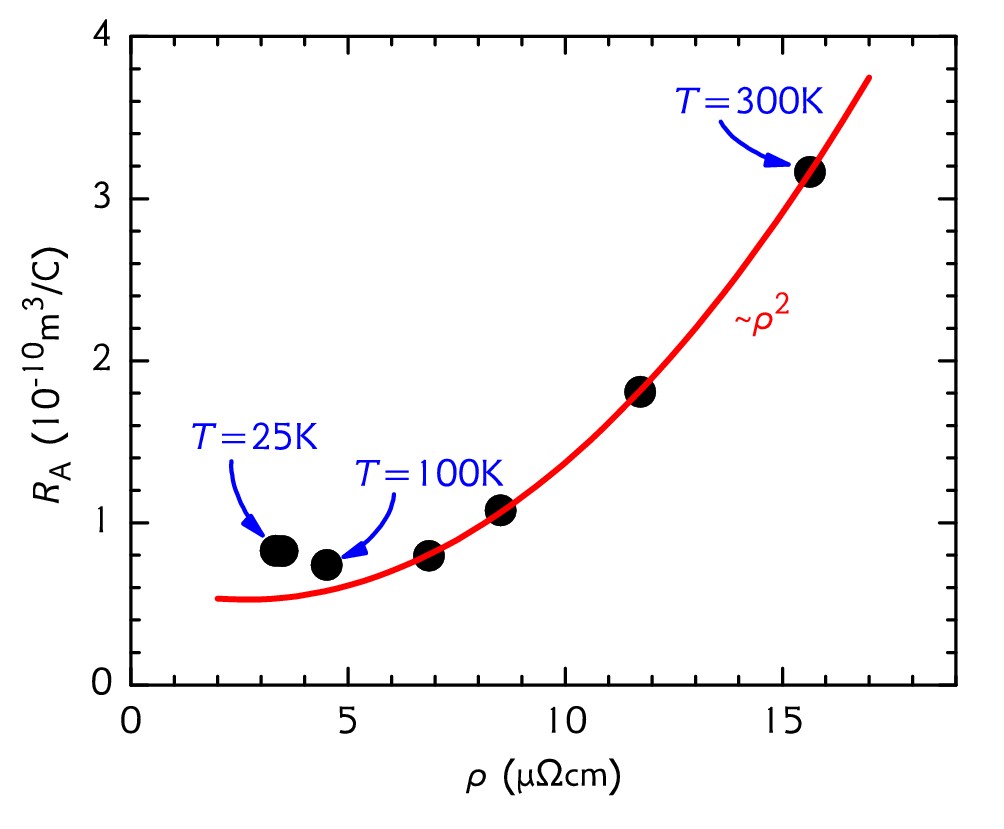} \centering
		\caption[$R_\text{A}(\rho_0)$]{Anomalous Hall coefficient $R_\text{A}$ in dependence of the zero field resistivity $\rho$ (black circles). Line: quadratic fit for temperatures above $100\,\text{K}$.}
		\label{fig:anohall}
	\end{figure}

	Thus, from our Hall data in high magnetic fields the ordinary Hall coefficient $R$ can be obtained, and with the knowledge of this coefficient and $\chi=5.4$, the anomalous Hall coefficient $R_\text{A}$ can be extracted from the low-field data, where $M = \chi \cdot H $. The results are shown in the inset of fig.~\ref{fig:hall}. The ordinary Hall coefficient $R$ is small and does not change much with temperature as expected for a metal. It slightly decreases with increasing temperature which indicates a change of the mobility of the charge carriers consistent with the multi-band nature of this material \cite{wurmehl2005}. At low temperatures the anomalous Hall coefficient $R_\text{A}$ is comparable to the ordinary one, while at $T\gtrsim 100\,\text{K}$ it strongly increases with $T$. The huge increase of $R_\text{A}$ cannot be explained by Lorentz force deflection, and implies the relevance of fundamental electronic deflection mechanisms known in ferromagnetic materials.  More specifically, following the classification given by Nagaosa et al. \cite{nagaosa2010}, one expects on the one hand $R_\text{A}\sim\rho^2$ for the \textit{intrinsic} Berry phase-related contribution to the AHE as well for the so-called \textit{side-jump} scattering contribution. On the other hand, for the extrinsic \textit{skew scattering}, $R_\text{A}\sim\rho$ is expected. In order to test the AHE in Co$_2$FeSi against one of these scenarios, we plot $R_\text{A}$ as a function of the measured $\rho(T)$ (fig.~\ref{fig:anohall}). Indeed, for $\rho$ larger than the residual resistivity $\rho_\text{R}$,	the data very well follow $R_\text{A}\propto (\rho-\rho_\text{R})^2+\mathrm{const}$, which provides clear evidence that the AHE is dominated by the intrinsic Berry phase and/or the side-jump scattering contributions. Interestingly, this resistivity regime corresponds to temperatures $T\gtrsim 100\,\text{K}$, i.e., to high temperatures where magnons are significantly populated, which may affect both the Berry phase as well as the side-jump scattering. At lower resistivity values, which correspond to the low-temperature regime where magnons apparently freeze out, the observed $\sim\rho^2$ behavior breaks down and $R_\text{A}$ becomes independent of $\rho$, implying that a qualitatively different deflection mechanism becomes dominant. Indeed the comparable magnitude of the anomalous ($R_\text{A}$) and ordiniary ($R$) Hall coefficients for $T\lesssim 100\,\text{K}$ suggests a contribution of the Lorentz force induced Hall effect (see inset fig.~\ref{fig:hall}), but also scew scattering contributions cannot be excluded. Note, that the temperature dependent change of $\rho$ in this regime is very small, and thus $R_\text{A}$ cannot be tested against the expected linear dependence of $\rho$. Impurity dependent investigations of $R_\text{A}$ should shed more light on this issue, but are beyond the scope of this study. It is, however, worth to mention that Imort et al., in agreement with the conjectured skew scattering, reported $R_\text{A}\sim\rho$ for Co$_2$FeSi thin films, which naturally contain more defects than our high quality crystal \cite{imort2011}. Finally, it is worth to point out that the easily accessible and by temperature well distinct regimes of the AHE, suggests Co$_2$FeSi as an ideal test bed for theoretical treatments of the AHE.
	
	In summary, we have investigated the electronic transport properties of Co$_2$FeSi in magnetic field. The temperature dependence of the resistivity is governed by electron-magnon scattering which exponentially freezes out over the substantial energy gap $\Delta\approx100\,\text{K}$ at low temperatures und thus provides clear-cut evidence of Co$_2$FeSi being a HMF.  This finding is corroborated by the magnetoresistance which is negative at high temperatures and positive at low temperatures. Furthermore, we observe that activation of ferromagnetic magnons at high temperatures $T\gtrsim\Delta$ has a strong impact on the anomalous Hall effect. In this high-temperature regime it is strongly temperature dependent and governed by Berry phase and side-jump deflection whereas at $T\lesssim\Delta$ it becomes temperature independent and is consistent with dominating skew scattering or intrinsic Lorentz force induced Hall effect.

\section*{Acknowledgements}
This work has been supported by the Deutsche Forschungsgemeinschaft through the Priority Programme SPP1538 (Grant No. HE 3439/9), and through the Emmy Noether Programme WU595/3-1 (S.W.).


\begin{thebibliography}{99}
	\bibitem{Chadov10}
		S. Chadov, X. Qi, J. K\"ubler, G. H. Fecher, C. Felser and S. C. Zhang, \emph{Tunable multifunctional topological insulators in ternary Heusler compounds}, Nat. Mat. {\bf 9}, 541 (2010)
	\bibitem{Liu12}
		J. Liu, T. Gottschall, K. P. Skokov, J. D. Moore and O. Gutfleisch, \emph{Giant magnetocaloric effect driven by structural transitions}, Nat. Mat. {\bf 11}, 620 (2012)
	\bibitem{Graf11}
		T. Graf, C. Felser and S.S.P. Parkin, \emph{Simple rules for the understanding of Heusler compounds}, Progress in Solid State Chemistry {\bf 39}, 1 (2011)
	\bibitem{degroot1983}
		R. A. de Groot, F. M. Mueller, P. G. van Engen and K. H. J. Buschow, \emph{New Class of Materials: Half-Metallic Ferromagnets}, Phys. Rev. Lett. {\bf 50}, 2024 (1983)
	\bibitem{mueller2009}
		G. M. M\"uller, J. Walowski, M. Djordjevic, G.-X. Miao, A. Gupta, A. V. Ramos, K. Gehrke, V. Moshnyaga, K. Samwer, J. Schmalhorst, A. Thomas, A. H\"utten, G. Reiss, J. S. Moodera and  M. M\"unzenberg, \emph{Spin polarization in half-metals probed by femtosecond spin excitation}, Nat. Mat. {\bf 8}, 56 (2009)
	\bibitem{shan2009}
		R. Shan, H. Sukegawa, W. H. Wang, M. Kodzuka, T. Furubayashi, T. Ohkubo,  S. Mitani, K. Inomata and K. Hono, \emph{Demonstration of Half-Metallicity in Fermi-Level-Tuned Heusler Alloy Co$_2$FeAl$_{0.5}$Si$_{0.5}$}, Phys. Rev. Lett. {\bf 102}, 246601 (2009)
	\bibitem{wurmehl2005}
		S. Wurmehl, G. H. Fecher, H. C. Kandpal, V. Ksenofontov, C. Felser, H.-J. Lin and J. Morais, \emph{Geometric, Electronic and Magnetic Structure of Co$_2$FeSi: Curie Temperature and Magnetic Moment Measurements and Calculations}, Phys. Rev. B {\bf 72}, 184434 (2005)
	\bibitem{blum}
		C. G. F. Blum, C. A. Jenkins, J. Barth, C. Felser, S. Wurmehl, G. Friemel, C. Hess, G. Behr, B. Büchner, A. Reller, S. Riegg, S. G. Ebbinghaus, T. Ellis, P. J. Jacobs, J. T. Kohlhepp and  H. J. M. Swagten, \emph{Highly ordered, half-metallic Co$_2$FeSi single crystaly}, Appl. Phys. Lett. \textbf{95}, 161903 (2009)
	\bibitem{wurmehl2006}
		S. Wurmehl, G. H. Fecher, H. C. Kandpal, V. Ksenofontov, C. Felser and H.-J. Lin, \emph{Investigations of Co$_2$FeSi: The Heusler Compound with Highest Curie Temperature and Magnetic Moment}, Appl. Phys. Lett. {\bf 88}, 032503 (2006)
	\bibitem{raquet2002}
		B. Raquet, M. Viret, E. Sondergard, O. Cespedes and R. Mamy, \emph{Electron-Magnon Scattering and Magnetic Resistivity in 3d Ferromagnets}, Phys. Rev. B {\bf 66}, 024433 (2002)
	\bibitem{raquet2002a}
		B. Raquet, M. Viret, J. M. Broto, E. Sondergard, O. Cespedes and R. Mamy, \emph{Magnetic Resistivity and Electron-Magnon Scattering in 3d Ferromagnets}, J. Appl. Phys. {\bf 91}, 8129 (2002)
	\bibitem{goodings1963}
		D. A. Goodings, \emph{Electrical Resistivity of Ferromagnetic Metals at Low Temperatues}, Phys. Rev. {\bf 132}, 542 (1963)
	\bibitem{isshiki1978}
		M. Isshiki and K. Igaki, \emph{Temperature Dependence of the Electrical Resistivity of Pure Iron at Low Temperatures}, Trans. JIM {\bf 19}, 431 (1978)
	\bibitem{isshiki1984}
		M. Isshiki, Y. Fukuda and K. Igaki, \emph{Temperature Dependence of the Electrical Resistivity of Pure Cobalt at Low Temperatures}, J. Phys. F {\bf 14}, 3007 (1984)
	\bibitem{otto1989}
		M. J. Otto, R. A. M. van Woerden, P. J. van der Valk, J. Wijngaard, C. F. van Bruggen and C. Haas, \emph{Half-Metallic Ferromagnets: II. Trabsport Properties of NiMnSb and Related Inter-Metallic Compounds}, J. Phys. Cond. Mat. {\bf 1}, 2351 (1989)
	\bibitem{nagaosa2010}
		N. Nagaosa, J. Sinova, S. Onoda, A. H. MacDonald and N. P. Ong, \emph{Anomalous Hall Effect}, Rev. Mod. Phys. {\bf 82}, 1539 (2010)
	\bibitem{imort2011}
		I.-M. Imort, P. Thomas, G. Reiss, and A. Thomas, \emph{Anomalous Hall effect in the Co-based Heusler compounds Co$_2$FeSi and Co$_2$FeAl},  J. Appl. Phys. \textbf{111}, 07D313 (2012)



\end{thebibliography}
\end{document}